\documentclass[aps,prd,amsmath,amssymb,superscriptaddress,preprintnumbers,twocolumn,10pt,nofootinbib]{revtex4-1}
\usepackage{verbatim}
\usepackage[T1]{fontenc}
\usepackage[utf8]{inputenc}
\usepackage[american]{babel}
\usepackage{enumitem}
\usepackage{transparent}
\usepackage{siunitx}
\usepackage{graphicx}
\usepackage{dcolumn}
\usepackage{bm}
\usepackage{amssymb}
\usepackage{latexsym}
\usepackage{booktabs}
\usepackage{amsmath}
\usepackage{multirow}
\usepackage{url}
\usepackage{footnote}
\usepackage{float}
\usepackage{threeparttable}
\usepackage{xcolor}
\definecolor{navyblue}{rgb}{0.0, 0.0, 0.5}
\definecolor{royalblue}{rgb}{0.25, 0.41, 0.88}
\definecolor{cadmiumgreen}{rgb}{0.0, 0.42, 0.24}
\definecolor{blue-violet}{rgb}{0.54, 0.17, 0.89}
\definecolor{darkviolet}{rgb}{0.58, 0.0, 0.83}
\definecolor{orange(colorwheel)}{rgb}{1.0, 0.5, 0.0}

\usepackage[normalem]{ulem}
\usepackage{color}
\usepackage{array}
\usepackage{enumerate}
\usepackage{adjustbox}

\usepackage{makecell}
\usepackage{diagbox}
\usepackage{epstopdf}
\usepackage{epsfig}
\usepackage{longtable}
\usepackage{supertabular}
\usepackage{algorithm}
\usepackage{pifont}
\usepackage{algorithmic}
\usepackage{changepage}
\usepackage{setspace}

\usepackage{hyperref}
\hypersetup{
    colorlinks=true,
    linkcolor=royalblue,
    citecolor=magenta
}

\begin{document}
    
\title{A Positive Turn: Coupled Quintessence and Cosmological Neutrino Masses}

\author{Tian-Nuo Li}\email{litiannuo@ucas.ac.cn}
\affiliation{School of Fundamental Physics and Mathematical Sciences, Hangzhou Institute for Advanced Study, UCAS, Hangzhou 310024, China}

\author{William Giar\`e}\email{giare@hawaii.edu}
\affiliation{Department of Physics and Astronomy, University of Hawai‘i, Honolulu, HI 96822, USA}

\author{Guo-Hong Du}\email{duguohong@stumail.neu.edu.cn}
\affiliation{Liaoning Key Laboratory of Cosmology and Astrophysics, College of Sciences, Northeastern University, Shenyang 110819, China}

\author{Eleonora Di Valentino}\email{e.divalentino@sheffield.ac.uk}
\affiliation{School of Mathematical and Physical Sciences, University of Sheffield, Hounsfield Road, Sheffield S3 7RH, United Kingdom}

\author{Luis A. Anchordoqui}\email{luis.anchordoqui@gmail.com}
\affiliation{Department of Physics and Astronomy, Lehman College, City University of New York, NY 10468, USA}

\author{\\Shao-Jiang Wang}\email{schwang@itp.ac.cn}
\affiliation{Institute of Theoretical Physics, Chinese Academy of Sciences (CAS), Beijing 100190, China}
\affiliation{Asia Pacific Center for Theoretical Physics (APCTP), Pohang 37673, Korea}

\author{Xin Zhang}\thanks{Corresponding author}\email{zhangxin@neu.edu.cn}
\affiliation{Liaoning Key Laboratory of Cosmology and Astrophysics, College of Sciences, Northeastern University, Shenyang 110819, China}
\affiliation{MOE Key Laboratory of Data Analytics and Optimization for Smart Industry, Northeastern University, Shenyang 110819, China}
\affiliation{National Frontiers Science Center for Industrial Intelligence and Systems Optimization, Northeastern University, Shenyang 110819, China}

\author{Rong-Gen Cai}\thanks{Corresponding author}\email{caironggen@nbu.edu.cn}
\affiliation{Institute of Fundamental Physics and Quantum Technology \& School of Physical Science and Technology, Ningbo University, Ningbo 315211, China}
\affiliation{Zhejiang Key Lab of Extreme Universe, Ningbo University, Ningbo 315211, China}
\affiliation{School of Fundamental Physics and Mathematical Sciences, Hangzhou Institute for Advanced Study, UCAS, Hangzhou 310024, China}

\begin{abstract}
\noindent In $\Lambda$CDM, the CMB and DESI BAO upper limits on the total neutrino mass are in tension with the lower bounds from oscillation experiments, and analyses allowing negative effective masses favor negative values. We investigate whether a coupled quintessence (CQ) model in which dark matter interacts with a canonical scalar field $\phi$ through an exponential mass coupling $m_{\rm c}(\phi)\propto e^{-\beta\phi/M_{\rm Pl}}$ can alleviate this tension. Combining CMB, DESI, and supernova data, we find that CQ yields neutrino-mass constraints comparable to those obtained in CPL and consistent with oscillation experiments within $2\sigma$ for normal ordering, without requiring the scalar-field equation of state to cross the phantom divide. We further show that it alleviates the tension more effectively than alternative interacting dark energy models that instead favor more negative effective masses. This comparison shows that whether dark sector interactions can relax neutrino mass bounds ultimately depends on the interplay of their effects on primary CMB anisotropies, geometry, growth, and lensing. \looseness=-1
\end{abstract}

\maketitle

\section{Introduction}
\label{sec1}

Baryon acoustic oscillations (BAO) from the Dark Energy Spectroscopic Instrument (DESI) provide the most precise determination of cosmological distances relative to the sound horizon at the baryon drag epoch, $r_{\rm d}$~\cite{DESI:2025zgx}. In a flat $\Lambda$CDM cosmology, these measurements primarily constrain the present-day matter fraction, $\Omega_{\rm m}$, and the combination $r_{\rm d} h$~\cite{DESI:2025zgx}, favoring lower $\Omega_{\rm m}$ and a higher $r_{\rm d} h$
than cosmic microwave background (CMB) data~\cite{Planck:2018nkj,Planck:2018vyg,ACT:2025fju,DESI:2025zgx,SPT-3G:2025bzu}. 

This $2$--$3\sigma$ discrepancy between the two probes~\cite{DESI:2025zgx,DESI:2025gwf,Ye:2025ark} directly affects the inference of cosmological neutrino mass bounds~\cite{Du:2024pai,Jiang:2024viw,Loverde:2024nfi,Capozzi:2025wyn,Lynch:2025ine,RoyChoudhury:2025dhe,Jhaveri:2025neg,Du:2025xes,Sharma:2025iux}. The most constraining limits on $\sum m_{\nu}$ combine information from the primary CMB anisotropies, the growth of structure probed by CMB lensing and the late-time geometry measured by BAO. At fixed physical baryon and cold dark matter (CDM) densities, increasing the neutrino mass raises the expansion rate during matter domination so that preserving the CMB acoustic scale requires a compensating reduction in the late-time expansion rate. In $\Lambda$CDM, this favors a higher $\Omega_{\rm m}$ and a lower
$r_{\rm d} h$, shifting the CMB constraints away from the region favored by DESI and driving the joint fit toward smaller masses~\cite{Jiang:2024viw,Loverde:2024nfi,DESI:2025zgx}. 

Additionally, larger neutrino masses reduce the lensing power by suppressing structure formation. A preference for enhanced lensing can therefore reinforce the geometric pull toward smaller masses.\footnote{Since the small-scale primary CMB tightly constrains $A_s e^{-2\tau}$, a larger reionization optical depth, $\tau$, implies a larger primordial amplitude and hence more lensing. This allows greater suppression by massive neutrinos, potentially relaxing the neutrino-mass bounds~\cite{Giare:2023ejv,Sailer:2025lxj,Jhaveri:2025neg}.} The well-known preference for excess lensing-induced smoothing of the acoustic peaks in Planck temperature and polarization data~\cite{Planck:2018vyg}, whose significance depends on the likelihood implementation~\cite{Rosenberg:2022sdy,Tristram:2023haj,Specogna:2024euz}, can therefore contribute to this tension~\cite{Capozzi:2025wyn}. More generally, allowing departures from the standard growth of matter perturbations can accommodate larger neutrino masses without modifying the background expansion~\cite{Giare:2025ath}.

In $\Lambda$CDM, the upper bounds on the total neutrino mass inferred from combinations of DESI BAO and CMB data are in substantial tension with the lower bound $\sum m_\nu \gtrsim 0.1$~eV required by oscillation experiments for inverted ordering (IO) and (depending on the precise CMB and lensing datasets) lie close to or even below the lower bound $\sum m_\nu \gtrsim 0.059$~eV allowed within the normal ordering (NO)~\cite{Jiang:2024viw,Esteban:2024eli,DESI:2025zgx,Elbers:2025vlz}. Moreover, when the same analyses are extended to an effective mass parameter that admits negative values, the posterior favors the negative region, yielding a $\sim3\sigma$ discrepancy with the NO lower bound~\cite{Craig:2024tky,Green:2024xbb,Elbers:2024sha,Naredo-Tuero:2024sgf,Elbers:2025vlz,Graham:2025dqn}.
Beyond $\Lambda$CDM, the tension between DESI BAO and CMB can be alleviated by dynamical dark energy (DE) models~\cite{Elbers:2024sha,Loverde:2024nfi,DESI:2025zgx,DESI:2025fii,Pulido-Hernandez:2026hcs,Yang:2026yaq,Kibris:2026cqq,Giare:2026oti}. Adopting the Chevallier--Polarski--Linder (CPL) parametrization of the DE equation of state~\cite{Chevallier:2000qy,Linder:2002et} yields a preference for evolving DE at the $\sim 3\sigma$ level~\cite{DESI:2025zgx,DESI:2025fii}. The additional freedom in the late-time expansion history allows the CMB acoustic scale and BAO distances to be reconciled at larger neutrino masses, relaxing the inferred bounds and restoring consistency with oscillation constraints~\cite{Elbers:2024sha,Elbers:2025vlz,Pulido-Hernandez:2026hcs,Yang:2026yaq}.

Within the CPL parametrization, however, the preferred DE evolution crosses the phantom divide from $w<-1$ at earlier times to $w>-1$ today~\cite{Giare:2024gpk,Cai:2025mas,Ozulker:2025ehg,Li:2025vuh,Li:2026asg,Zhang:2026zbs}. This behavior cannot be realized by a single minimally
coupled canonical scalar field and calls for a well-defined physical realization beyond phenomenological parametrizations~\cite{Feng:2004ff,Vikman:2004dc}.\footnote{Proposed explanations of the phantom-crossing behaviour suggested by DESI include (but are not limited to) two-field quintom models~\cite{Feng:2004ad,Guo:2004fq,Zhang:2005kj,Goh:2025upc}, higher-derivative scalar theories~\cite{Li:2005fm,Creminelli:2008wc}, non-minimally coupled gravity~\cite{Perivolaropoulos:2005yv,Chudaykin:2024gol,Ye:2024ywg,Pan:2025psn,Wolf:2025jed,SanchezLopez:2025uzw,Efstratiou:2025iqi}, kinetic-braiding and Galileon models~\cite{Deffayet:2010qz,Matsumoto:2017qil,Tiwari:2024gzo,Tsujikawa:2025wca,Cataneo:2025vae,Naidoo:2026umv}, scalar--vector--tensor theories~\cite{Roy:2026jap}, minimally modified gravity~\cite{Yang:2024kdo,Arora:2025msq}, braneworld scenarios~\cite{Sahni:2002dx,Mishra:2025goj,Mukherjee:2026lnb}, fermion condensation with cosmological backreaction~\cite{Liu:2026nds}, and vacuum phase transitions followed by decay~\cite{Giare:2026opg}. For reviews, see e.g., Refs.~\cite{Cai:2025mas,Li:2026asg,Anchordoqui:2026hys,Zhang:2026zbs}.}

In this regard, interactions between dark matter (DM) and DE provide an alternative explanation of the DESI results~\cite{Nunes:2016dlj,DiValentino:2019ffd,Mukhopadhyay:2020bml,Zhao:2022ycr,Mishra:2023ueo,vanderWesthuizen:2023hcl,Escamilla:2023shf,Castello:2023zjr,Halder:2024uao,Silva:2024ift,Giare:2024smz,Pooya:2024wsq,Li:2024qso,Sabogal:2024yha,Lewis:2024cqj,Li:2025owk,Sabogal:2025mkp,Tsedrik:2025cwc,Silva:2025hxw,You:2025uon,Pan:2025qwy,Yashiki:2025loj,Li:2025ula,Bedroya:2025fwh,Wang:2025znm,Yang:2025uyv,vanderWesthuizen:2025rip,Li:2025muv,Zhang:2025dwu,Zhang:2025lam,Figueruelo:2026eis,Escobal:2026zxb,Li:2026xaz,Dai:2026pvx,Kashyap:2026ivg,Kolhatkar:2026ixl,Wang:2026wrk,Wang:2026vqw,Zhai:2026uwr,Li:2026asg,Zhang:2026dtd,Yang:2026qmz,Sabogal:2026bsw,Babu:2026tfr,Bagherian:2026jxz,Li:2026jgg,Wang:2026crv}. At the background level, energy exchange within the dark sector can modify the expansion history similarly to an evolving DE equation of state~\cite{vonMarttens:2019ixw,Sabogal:2026bsw}. Unlike non-interacting models such as CPL, however, interactions also change the dilution of the DM density, which no longer scales as $a^{-3}$. The inferred phantom behaviour of DE can thus be a mirage of dark sector interactions when the expansion history is interpreted assuming separately conserved DM and DE components~\cite{Das:2005yj,vonMarttens:2019ixw,Chakraborty:2025syu,Khoury:2025txd,Guedezounme:2025wav,Chen:2025ywv,Liu:2025bss,Li:2026xaz,Antusch:2026ldp,Khoury:2026svx,Ganguly:2026mgu}. For an explicit theoretical realization, see Appendix~\ref{app:DD}.

It was recently shown that coupled quintessence (CQ), in which DE is described by a scalar field coupled to DM through a well-defined Lagrangian interaction, provides a fit to observations competitive with CPL, without requiring the scalar field equation of state to cross the phantom divide~\cite{Bedroya:2025fwh,Li:2026xaz}. In the CQ regime favored by the data, energy flows from DM to DE, reducing the late-time DM abundance relative to the $a^{-3}$ extrapolation of its early-time density~\cite{Li:2026xaz}. The coupling also modifies the DM perturbation dynamics, the growth of structure and CMB lensing, thus directly affecting the relevant contributions underlying neutrino-mass inference~\cite{Li:2026xaz}. 

In this work, we show that these effects alleviate the tension between cosmological neutrino mass bounds and oscillation experiments, allowing CQ to accommodate masses consistent with the NO within $\sim 2\sigma$. We further show that CQ alleviates such tension more effectively than some alternative interacting models further establishing this scenario as a competitive alternative to CPL.\footnote{Recently, Ref.~\cite{Wang:2026gvg} also investigated an exponentially coupled quintessence model with a freely varying effective neutrino mass. Here we consider a different quintessence potential and include additional data from small-scale CMB experiments and several supernova samples to assess the tension between cosmological neutrino mass constraints and oscillation experiments, using both marginalized posteriors and profile likelihoods.}

The paper is organized as follows. In Section~\ref{sec2}, we introduce the CQ model and the effective neutrino mass prescription. In Section~\ref{sec3}, we describe our methodology and datasets. We present our results in Section~\ref{sec4} and conclude in Section~\ref{sec5}. Appendix~\ref{app:DD} discusses the theoretical motivation, Appendix~\ref{app:SN} examines alternative supernova samples, and Appendix~\ref{app:IDE} presents additional results for phenomenological interacting dark energy (IDE) models.

\section{Coupled Quintessence and Effective Neutrino Mass}\label{sec2}

In the CQ model, DE is described by a canonical scalar field $\phi$ that exchanges energy and momentum with CDM~\cite{Amendola:1999er,Pettorino:2008ez,Li:2026xaz}. Following the convention of Ref.~\cite{Li:2023fdk}, the energy-momentum transfer is described by
\begin{equation}
 \nabla_\nu T^\nu{}_{\mu,{\rm de}}
 =Q_\mu,\qquad \nabla_\nu T^\nu{}_{\mu,{\rm c}}=-Q_\mu.
 \label{eq:cq-transfer}
\end{equation}
Here $T^\nu{}_{\mu,I}$ is the energy-momentum tensor of component $I$, and $Q_\mu$ is the transfer vector for DE. The opposite signs ensure conservation of the combined dark sector. At the background level,
\begin{equation}
 \begin{aligned}
 \rho_{\rm de}'+3\mathcal H(1+w)\rho_{\rm de}&=aQ,\\
 \rho_{\rm c}'+3\mathcal H\rho_{\rm c}&=-aQ ,
 \end{aligned}
\end{equation}
where $\rho_{\rm de}$ and $\rho_{\rm c}$ are the energy densities of DE and CDM, respectively, while $w=p_{\rm de}/\rho_{\rm de}$ is the DE equation-of-state parameter, with pressure $p_{\rm de}$. Primes denote conformal-time derivatives, and $\mathcal H=a'/a$. The background energy transfer rate $Q$ is defined by $Q_0=-aQ$.

For the exponential coupling, the transfer vector is
\begin{equation}
 Q_\mu=-\frac{\beta\rho_{\rm c}}{M_{\rm Pl}}\,\partial_\mu\phi ,
 \label{eq:cq-vector}
\end{equation}
where $\beta$ is the dimensionless coupling constant, $M_{\rm Pl}=(8\pi G)^{-1/2}$ is the reduced Planck mass with $G$ Newton's constant, and $\partial_\mu\phi$ is the scalar-field gradient. Its time component gives
\begin{equation}
 Q=\frac{\beta\rho_{\rm c}\phi'}{aM_{\rm Pl}}.
\end{equation}
For $Q>0$, energy is transferred from CDM to DE; for $Q<0$, the direction is reversed.

We adopt an inverse power-law potential,
\begin{equation}
 U(\phi)=U_0\left(\frac{\phi}{M_{\rm Pl}}\right)^{-\alpha},
\end{equation}
where $U_0$ is the potential normalization and $\alpha$ controls its slope. The canonical field has density $\rho_{\rm de}=\phi'^2/(2a^2)+U$ and pressure $p_{\rm de}=\phi'^2/(2a^2)-U$, so $w$ follows from the field evolution.\footnote{The potential's shape (inverse power-law vs. exponential) has little impact on the fit to current cosmological data~\cite{Anchordoqui:2026hys}.}

We allow the effective neutrino mass sum $\sum m_{\nu,\mathrm{eff}}$ to vary across zero. For a species with comoving momentum magnitude $q$, we replace the usual comoving energy by~\cite{Elbers:2024sha,Elbers:2025vlz}
\begin{equation}
 \epsilon_{\rm eff}=s\sqrt{q^2+a^2m_{\nu,\mathrm{eff}}^2},
 \qquad s=\operatorname{sgn}(m_{\nu,\mathrm{eff}}).
\end{equation}
The usual massive-neutrino equations are recovered for positive masses. At zero mass we use the standard massless solution. A negative value labels the reversed gravitational contribution of the massive species; it does not denote a negative particle rest mass.

For nonzero mass, let $\Psi_\ell$ be the $\ell$th multipole of the neutrino phase-space perturbation and define
\begin{equation}
 \widetilde{\Psi}_\ell=s^\ell\Psi_\ell .
\end{equation}
The factors of $s$ on neighboring multipoles cancel the sign in the streaming coefficient $q/\epsilon_{\rm eff}$. In these variables the hierarchy takes the positive-energy form, while the stress-energy sources are
\begin{equation}
 \left(\delta\rho_\nu,\delta p_\nu,J_\nu,\Pi_\nu\right)_{\rm eff}
 =s\left(\delta\rho_\nu,\delta p_\nu,J_\nu,\Pi_\nu\right)_{+}.
\end{equation}
Here $\delta\rho_\nu$, $\delta p_\nu$, $J_\nu$, and $\Pi_\nu$ are the density and pressure perturbations, momentum density, and anisotropic stress of the massive sector. The subscript $+$ denotes integrals using $|\epsilon_{\rm eff}|$ and $\widetilde{\Psi}_\ell$ evaluated in the metric of the signed model. The background density and pressure of the massive sector also carry the factor $s$.

On the negative branch, reversing the massive contribution removes twice its original relativistic density. We therefore increase the massless normalization from $N_{\rm ur}$ to $N_{\rm ur}+2N_{\rm m}$, where $N_{\rm m}$ denotes the relativistic normalization of the massive species before the sign change. This keeps the total at $N_{\rm eff}=3.044$ in the relativistic limit~\cite{Elbers:2025vlz}.

\section{Methodology and Data}\label{sec3}

We implement both the CQ model and the effective neutrino mass prescription in a modified version of \texttt{CAMB}~\cite{Lewis:1999bs} and perform Markov Chain Monte Carlo analyses using the \texttt{Cobaya} sampler~\cite{Torrado:2020dgo}. We vary the standard six $\Lambda$CDM parameters, the two additional DE parameters $(\alpha,\beta)$ in CQ and the effective neutrino mass sum. To compare the impact of CQ on the effective neutrino mass posteriors, we separately run $\Lambda$CDM+$\sum m_{\nu, \rm eff}$ and CPL+$\sum m_{\nu, \rm eff}$. For the latter, we assume a linearly evolving DE equation of state $w(a)=w_0+w_a(1-a)$~\cite{Chevallier:2000qy,Linder:2002et}.  

The flat prior ranges on all sampled parameters are given in Table~\ref{tab:parameter_priors}. We assess convergence using the Gelman--Rubin statistic, requiring $R-1<0.02$~\cite{Gelman:1992zz}.

\begin{table}[t]
\centering
\renewcommand{\arraystretch}{1.2}
\begin{tabular}{l@{\hspace{2cm}}c}
\hline\hline
\textbf{Parameter} & \textbf{Prior} \\
\hline
$\Omega_{\rm b} h^2$ & $[0.005,\,0.1]$ \\
$\Omega_{\rm c} h^2$ & $[0.01,\,0.99]$ \\
$H_0$~[km/s/Mpc] & $[20,\,100]$ \\
$\tau$ & $[0.01,\,0.8]$ \\
$\ln(10^{10}A_{\rm s})$ & $[1.61,\,3.91]$ \\
$n_{\rm s}$ & $[0.8,\,1.2]$ \\
$\sum m_{\nu,\mathrm{eff}}$~[eV] & $[-5,\,5]$ \\
$\alpha$ & $[0,\,1.4]$ \\
$\beta$ & $[0,\,0.15]$ \\
$w_0$ & $[-3,\,1]$ \\
$w_a$ & $[-3,\,2]$ \\
\hline\hline
\end{tabular}
\caption{Flat priors on the sampled cosmological parameters.}
\label{tab:parameter_priors}
\end{table}

The datasets employed in our analysis are detailed below:
\begin{itemize}
   \item \textbf{CMB}: The CMB temperature and polarization power spectra from Planck, ACT DR6, and SPT-3G are combined in our analysis. For Planck, we use the low-$\ell$ \texttt{Commander} and \texttt{SimAll} likelihoods, together with the intermediate-$\ell$ \texttt{CamSpec} likelihood~\cite{Efstathiou:2019mdh,Rosenberg:2022sdy}. The high-$\ell$ ACT and SPT-3G measurements are included through the \texttt{ACT-lite}~\cite{ACT:2025fju} and \texttt{SPT-lite} likelihoods~\cite{SPT-3G:2025bzu}, respectively, with the multipole ranges chosen to avoid double counting. We also include the joint CMB lensing power-spectrum likelihood combining reconstructions from the Planck, ACT, and SPT-3G maps~\cite{Carron:2022eyg,ACT:2023ubw,ACT:2023dou,ACT:2023kun,SPT-3G:2024atg,ACT:2025qjh}.
   \item \textbf{DESI}: The BAO measurements obtained from DESI DR2 using galaxies, quasars, and the Lyman-$\alpha$ forest are reported in Ref.~\cite{DESI:2025zgx}.
   \item \textbf{DES-Dovekie}: The DES-Dovekie dataset is a recalibrated and reanalyzed Type Ia supernova sample based on the DES five-year survey and the Dovekie cross-calibration program, comprising approximately 1800 supernovae~\cite{DES:2025sig}.
\end{itemize}

\section{Results}\label{sec4}

\begin{table*}[t!]
\centering
\renewcommand{\arraystretch}{1.5}
\resizebox{0.95\textwidth}{!}{%
\begin{tabular}{l@{\hspace{0.5cm}}ccc@{\hspace{1cm}}ccc}
\toprule
&
\multicolumn{3}{c}{CMB+DESI}
&
\multicolumn{3}{c}{CMB+DESI+DES-Dovekie}
\\
\cmidrule(lr){2-4}
\cmidrule(lr){5-7}
Parameter
& $\Lambda$CDM+$\sum m_{\nu,\mathrm{eff}}$
& CPL+$\sum m_{\nu,\mathrm{eff}}$
& CQ+$\sum m_{\nu,\mathrm{eff}}$
& $\Lambda$CDM+$\sum m_{\nu,\mathrm{eff}}$
& CPL+$\sum m_{\nu,\mathrm{eff}}$
& CQ+$\sum m_{\nu,\mathrm{eff}}$ 
\\
\midrule

$\sum m_{\nu,\mathrm{eff}}\ [{\rm eV}]$
& $-0.051^{+0.032}_{-0.041}$
& $0.024^{+0.054}_{-0.064}$
& $-0.007^{+0.038}_{-0.070}$
& $-0.042^{+0.033}_{-0.040}$
& $-0.005^{+0.043}_{-0.059}$
& $-0.017^{+0.041}_{-0.063}$
\\
$P(\sum m_{\nu,\mathrm{eff}}<0)\ [\%]$
& $90.0$
& $36.6$
& $61.5$
& $86.7$
& $56.3$
& $66.7$
\\
$P(\sum m_{\nu,\mathrm{eff}}<0.06\,{\rm eV})\ [\%]$
& $99.65$
& $73.6$
& $86.4$
& $99.47$
& $88.5$
& $90.5$
\\
$P(\sum m_{\nu,\mathrm{eff}}<0.10\,{\rm eV})\ [\%]$
& $>99.9$
& $89.8$
& $94.6$
& $>99.9$
& $96.8$
& $97.0$
\\
\addlinespace

\hline
$w_0$
& --
& $-0.480\pm0.220$
& --
& --
& $-0.825\pm0.057$
& --
\\

$w_a$
& --
& $-1.52^{+0.74}_{-0.65}$
& --
& --
& $-0.59^{+0.26}_{-0.21}$
& --
\\

$\beta$
& --
& --
& $0.038^{+0.016}_{-0.013}$
& --
& --
& $0.041^{+0.015}_{-0.012}$
\\

$\alpha$
& --
& --
& $<0.624$
& --
& --
& $0.39^{+0.17}_{-0.19}$
\\
\addlinespace

\hline
$H_0\ [{\rm km\,s^{-1}\,Mpc^{-1}}]$
& $68.79\pm0.32$
& $64.30^{+1.80}_{-2.10}$
& $68.43^{+0.82}_{-0.59}$
& $68.62\pm0.31$
& $67.53\pm0.55$
& $68.00\pm0.57$
\\

$\Omega_{\rm m}$
& $0.2973^{+0.0036}_{-0.0041}$
& $0.3460\pm0.0220$
& $0.2970^{+0.0055}_{-0.0063}$
& $0.2994\pm0.0038$
& $0.3109\pm0.0056$
& $0.2994\pm0.0052$
\\

$\sigma_8$
& $0.8341\pm0.0081$
& $0.7920\pm0.0200$
& $0.8344\pm0.0099$
& $0.8324\pm0.0083$
& $0.8218\pm0.0091$
& $0.8322\pm0.0096$
\\

$S_8$
& $0.8303\pm0.0078$
& $0.8490\pm0.0110$
& $0.8301\pm0.0079$
& $0.8315\pm0.0078$
& $0.8364\pm0.0078$
& $0.8313\pm0.0082$
\\
\addlinespace

\hline
$\Delta\chi^2_{\rm MAP,CMB}$
& $0$
& $-1.79$
& $-0.77$
& $0$
& $-2.18$
& $-0.08$
\\

$\Delta\chi^2_{\rm MAP,BAO}$
& $0$
& $-2.65$
& $-0.99$
& $0$
& $-1.80$
& $-1.61$
\\

$\Delta\chi^2_{\rm MAP,SN}$
& --
& --
& --
& $0$
& $-6.35$
& $-4.29$
\\

$\Delta\chi^2_{\rm MAP,total}$
& $0$
& $-4.45$
& $-1.76$
& $0$
& $-10.33$
& $-5.98$
\\

\bottomrule
\end{tabular}%
}
\caption{Main results for the $\Lambda$CDM, CPL, and CQ models with a free effective neutrino mass sum, for the two dataset combinations considered. The first block reports constraints on $\sum m_{\nu,\mathrm{eff}}$ and posterior probabilities below zero and below the NO and IO benchmarks of $0.06$ and $0.10\,\mathrm{eV}$, respectively. The second block gives constraints on the DE parameters in CPL and CQ, while the third reports constraints on some informative cosmological parameters. Uncertainties are given at $68\%$ confidence level (CL), while upper limits are given at $95\%$ CL. The final block lists the $\Delta\chi^2_{\rm MAP}$ contributions relative to $\Lambda$CDM+$\sum m_{\nu,\mathrm{eff}}$ for the same data combination; negative values indicate an improved fit.}
\label{tab:results}
\end{table*}

\begin{figure}[!htp]
\centering
\includegraphics[width=0.45\textwidth]{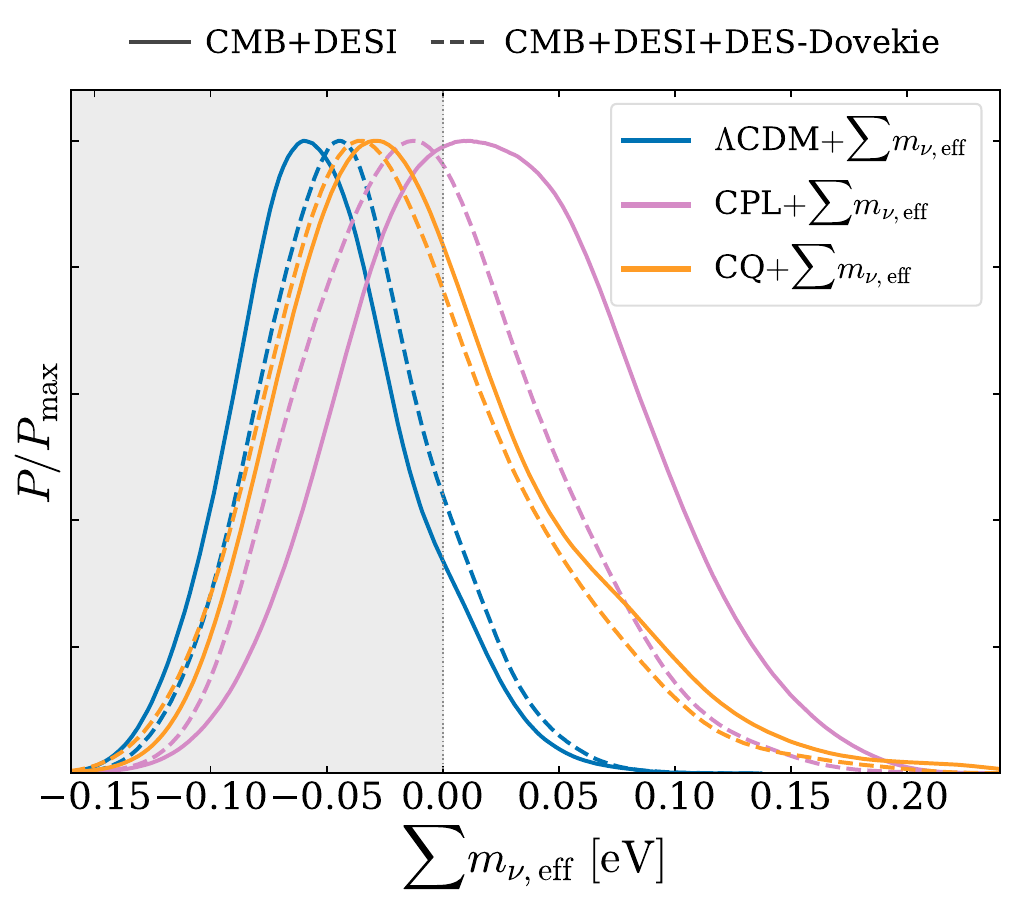}
\caption{Marginalized posterior distributions of $\sum m_{\nu,\mathrm{eff}}$ for the $\Lambda$CDM+$\sum m_{\nu,\mathrm{eff}}$, CPL+$\sum m_{\nu,\mathrm{eff}}$, and CQ+$\sum m_{\nu,\mathrm{eff}}$ models, obtained from CMB+DESI and CMB+DESI+DES-Dovekie.}
\label{fig:CQ_mnu:posteriors}
\end{figure}

\begin{figure}[!htp]
\centering
\includegraphics[width=0.48\textwidth]{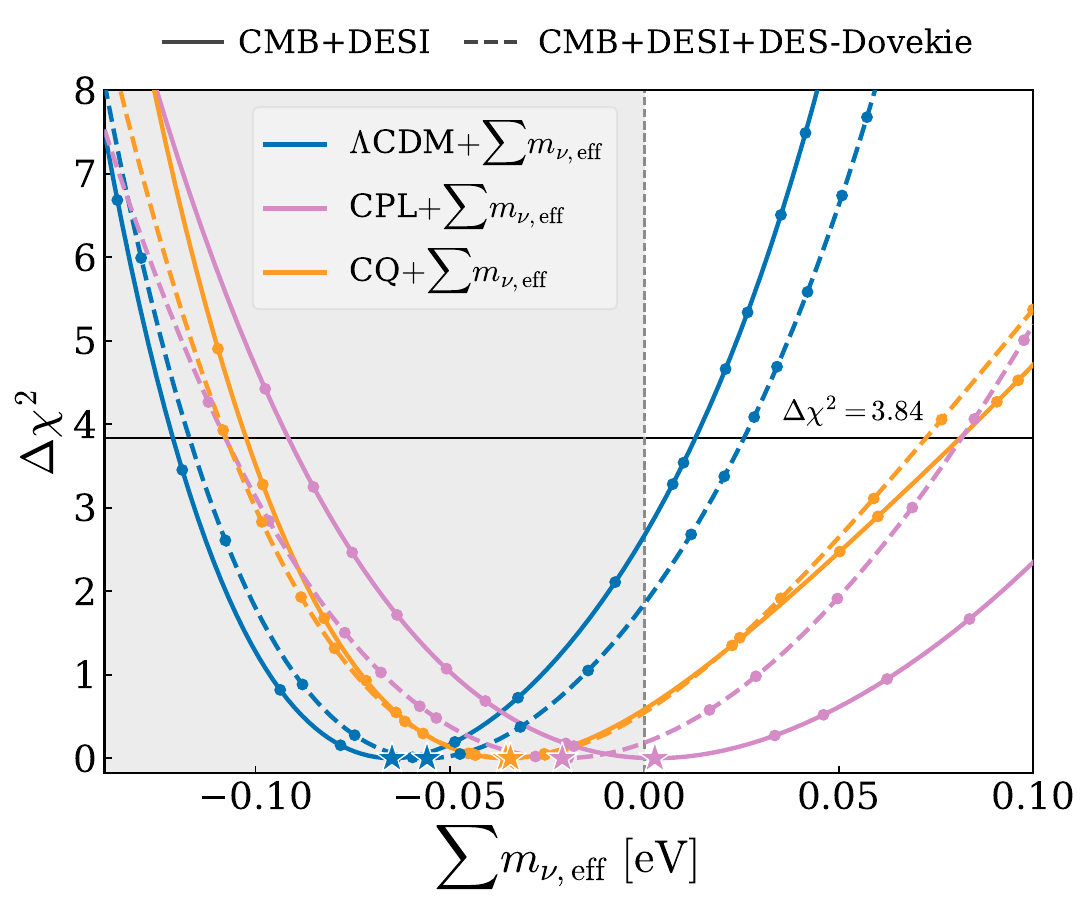}
\caption{
Profile likelihoods of $\sum m_{\nu,\mathrm{eff}}$ for the
$\Lambda$CDM+$\sum m_{\nu,\mathrm{eff}}$, CPL+$\sum m_{\nu,\mathrm{eff}}$, and CQ+$\sum m_{\nu,\mathrm{eff}}$ models, obtained from CMB+DESI and CMB+DESI+DES-Dovekie. The stars mark the best-fit values determined from the profile likelihoods.}
\label{fig:CQ_mnu:profile}
\end{figure}

\begin{figure*}[!htp]
\centering
\includegraphics[width=1\textwidth]{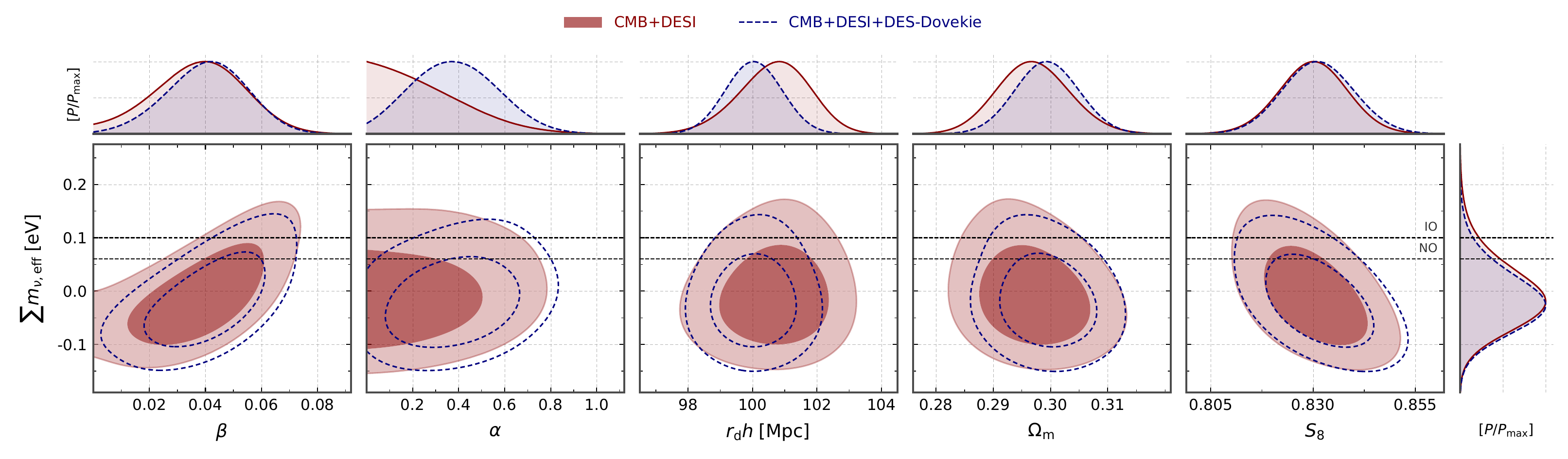}
\caption{A comparison of the two-dimensional marginalized contours at 68\% and 95\% CL between $\sum m_{\nu,\mathrm{eff}}$ and the cosmological parameters
$\beta$, $\alpha$, $r_d\,h$, $\Omega_{\rm m}$, and $S_8$ for the CQ+$\sum m_{\nu,\mathrm{eff}}$ model, using CMB+DESI and CMB+DESI+DES-Dovekie.}
\label{fig:CQ_mnu:params}
\end{figure*}

Table~\ref{tab:results} summarizes the main results for all models, while Fig.~\ref{fig:CQ_mnu:posteriors} shows the one-dimensional marginalized posterior distributions of the total effective neutrino mass obtained from the CMB+DESI and CMB+DESI+DES-Dovekie data.

Within $\Lambda$CDM, both data combinations produce posterior distributions that peak at negative mass values, with $90.0\%$ and $86.7\%$ of the total probability lying at $\sum m_{\nu,\mathrm{eff}}<0$, respectively.

Allowing a CPL equation of state shifts the posterior toward larger masses. For CMB+DESI we get $\sum m_{\nu,\mathrm{eff}}=0.024^{+0.054}_{-0.064}\,\mathrm{eV}$, reducing the probability below zero to $36.6\%$. Adding DES-Dovekie, however, narrows the posterior and shifts its mean back below zero. This increases the probability of negative effective masses to $56.3\%$.

For CQ, we obtain $\sum m_{\nu,\mathrm{eff}}=-0.007^{+0.038}_{-0.070}\,\mathrm{eV}$ from CMB+DESI and $-0.017^{+0.041}_{-0.063}\,\mathrm{eV}$ after adding DES-Dovekie. Both values are consistent with zero at the $68\%$ CL. Relative to $\Lambda$CDM, the posterior distributions extend further into the positive-mass region, reducing the probabilities below zero to $61.5\%$ and $66.7\%$. These remain higher than in CPL, but change less upon adding DES-Dovekie. Overall, the CQ constraints remain similarly stable across alternative supernova samples, as examined in Appendix~\ref{app:SN}.

Regarding the neutrino mass ordering, CQ has $13.6\%$ and $9.5\%$ of the posterior above the NO benchmark of $0.06\,\mathrm{eV}$ for CMB+DESI and CMB+DESI+DES-Dovekie, respectively, compared with only $0.35\%$ and $0.53\%$ in $\Lambda$CDM. CPL has more probability above this threshold ($26.4\%$ and $11.5\%$). For the full CMB+DESI+DES-Dovekie combination, however, the difference between CPL and CQ is only $\sim2$ percentage points. In both models, the IO benchmark of $0.10\,\mathrm{eV}$ is less well accommodated, as only $\sim3\%$ of the posterior lies above it after adding DES-Dovekie.

The profile likelihoods in Fig.~\ref{fig:CQ_mnu:profile} support the same comparison. The $\Lambda$CDM profiles have minima at negative masses and rise steeply toward positive values, whereas CQ and CPL yield broader profiles, with $\Delta\chi^2<3.84$ at $\sum m_{\nu,\mathrm{eff}}=0.06\,\mathrm{eV}$ for both data combinations. The NO benchmark is therefore allowed by both the marginalized posteriors and the profile likelihoods in CQ, restoring consistency with oscillation experiments within $2\sigma$.

Figure~\ref{fig:CQ_mnu:params} shows the joint CQ constraints on the effective neutrino mass and selected cosmological parameters. Larger masses correlate with larger coupling strengths, $\beta$, and smaller values of $S_8$. The coupling itself changes little between the two data combinations, with $\beta=0.038^{+0.016}_{-0.013}$ for CMB+DESI and $0.041^{+0.015}_{-0.012}$ after adding DES-Dovekie. The supernova data change the constraint on the potential parameter $\alpha$ from an upper limit to a two-sided interval, while the marginalized constraints on $\Omega_{\rm m}$ and $S_8$ remain close to their $\Lambda$CDM values.

Both extensions also improve the fit relative to $\Lambda$CDM+$\sum m_{\nu,\mathrm{eff}}$, as reported in Table~\ref{tab:results}. For CMB+DESI, CQ and CPL give $\Delta\chi^2_{\rm MAP}=-1.76$ and $-4.45$, respectively. Adding DES-Dovekie, these become $-5.98$ and $-10.33$. CPL thus provides the larger improvement for both combinations. For CQ with DES-Dovekie, the improvement comes predominantly from supernovae, followed by BAO, with little change in the CMB contribution. 

Additional interacting DE models are examined in Appendix~\ref{app:IDE}. Their effective neutrino-mass posteriors have more negative central values than CQ for both baseline data combinations, showing that the shifts found in CQ are not shared equally by the other interacting scenarios considered here.

\section{Conclusion and discussion}\label{sec5}

We have shown that CQ alleviates the tension between cosmological neutrino-mass constraints and oscillation experiments, accommodating the NO benchmark within $2\sigma$ in both the marginalized posteriors and the profile likelihoods. Although CPL allows more posterior probability above the NO benchmark, the two models give similar probabilities once DES-Dovekie is included. CQ thus provides a competitive alternative to CPL without requiring the scalar-field equation of state to cross the phantom divide.

In CQ, larger neutrino masses are accommodated at stronger coupling, see Fig.~\ref{fig:CQ_mnu:params}. Stronger DM to DE energy transfer reduces the late-time CDM abundance, leaving room for a larger neutrino contribution to the total matter density. This also provides a connection with studies showing that a reduced DE contribution to the expansion rate at intermediate redshifts can accommodate larger neutrino masses~\cite{Yang:2026yaq,Kibris:2026cqq}. In CQ, assuming separately conserved CDM with the same early-time normalization, $\rho_{\rm c}^{\rm ref}\propto a^{-3}$, leads to an inferred effective DE density $\rho_{\rm de}^{\rm eff}=\rho_\phi+\rho_{\rm c}-\rho_{\rm c}^{\rm ref}$. Although the scalar-field density is enhanced, the CDM deficit contributes negatively to the effective DE density. For the central CMB+DESI+DES-Dovekie parameters, $\rho_{\rm de}^{\rm eff}$ decreases with increasing redshift beyond $z\sim1$ and falls below the DE density inferred in $\Lambda$CDM at $z\sim2$. A direct comparison of the expansion histories shows that CQ also predicts a slightly lower $H(z)$ than $\Lambda$CDM at $z\gtrsim3$.

More generally, compensating the matter density does not preserve its clustering properties because neutrinos free-stream out of small-scale overdensities, whereas CDM clusters on those scales. For these coupling values, two effects therefore contribute to structure growth. On the one hand, increasing the neutrino mass suppresses growth, as reflected in its anticorrelation with $S_8$ in Fig.~\ref{fig:CQ_mnu:params}. On the other hand, the interaction itself modifies the dynamics of CDM perturbations. The resulting CMB lensing signal depends on both effects. The connection with a reduced effective contribution to the expansion history must therefore be assessed together with the perturbation dynamics.

Overall, addressing the neutrino-mass tension with DM-DE interactions requires that the changes induced in the late-time geometry, primary CMB anisotropies, clustering, and lensing combine constructively to relax the neutrino-mass limits.
Some alternative interacting models considered in Appendix~\ref{app:IDE} also suppress the DE contribution at intermediate redshifts and favor substantially larger matter fractions and lower clustering amplitudes, yet still show a preference for more negative effective neutrino masses. This comparison highlights that neither increasing $\Omega_{\rm m}$ nor lowering the clustering amplitude necessarily leads to weaker neutrino-mass constraints, see also Ref.~\cite{Yang:2026qon}. The models must also reconcile the early-time densities inferred from the CMB with the late-time distances measured by BAO, while reproducing the lensing signal. This provides a useful criterion for discriminating among competing models that offer comparably good fits to observations. In this sense, our results further establish CQ as a competitive explanation of the DESI results.

\section*{Acknowledgments}
We thank Yun-He Li, Cumrun Vafa, and David Wu for helpful discussions. This work was supported by the National Natural Science Foundation of China Grants (No. 12533001, No.12473001, No. 12422502, No. 12547110, No. 12588101, No. 12235019, and No. 12447101), the National SKA Program of China Grants (No. 2022SKA0110200 and No. 2022SKA0110203), the National Key Research and Development Program of China Grants (No. 2021YFC2203004 and No. 2021YFA0718304), and the 111 Project (Grant No. B16009). WG acknowledges support from the National Aeronautics and Space Administration (NASA) under Grant No.~80NSSC24K0898. EDV is supported by a Royal Society Dorothy Hodgkin Research Fellowship. The work of LAA is supported by the U.S. National Science Foundation (NSF Grant PHY-2412679), he extends his appreciation to the Harvard Swampland Initiative for their warm hospitality and for providing a stimulating environment for productive discussions.

\bibliography{main}

\newpage

\appendix

\section{Theoretical Motivation}
\label{app:DD}

The dark dimension has emerged as a compelling five-dimensional (5D) framework that unifies the macroscopic parameters of the dark sector into a singular geometric origin~\cite{Montero:2022prj,Gonzalo:2022jac}. Derived from the intersections of string theory and quantum gravity under the Swampland program~\cite{Vafa:2005ui}, this paradigm bridges a deep conceptual gap between high-energy theory and experiment. Rather than treating the smallness of the cosmological constant ($\Lambda \sim 10^{-120} M_{\rm Pl}^4$) as a fine-tuning anomaly, the Swampland distance conjecture~\cite{Ooguri:2006in} and anti-de Sitter distance conjecture~\cite{Lust:2019zwm} dictate that this small vacuum energy mandates a unique corner of the string landscape with a single, mesoscopic extra dimension. Crucially, this specific 5D compactification scale is not an arbitrary theoretical choice, but a tight constraint arising from the direct compatibility of Swampland criteria and empirical observation; most notably, precision tabletop tests of Newton’s inverse-square law~\cite{Lee:2020zjt,Tan:2020vpf}. To be more specific, the dark dimension is effectively modeled as a 5D flat spacetime, where four dimensions are non-compact, and the extra spatial dimension, $y$, is compactified on a circle $S^{1}$ of radius $R \sim 1~\mu{\rm m}$. The lack of chirality inherent to standard $S^{1}$ compactifications is circumvented through the implementation of an $S^1/\mathbb{Z}_2$ orbifold geometry. This configuration forces the bulk fields to transform under the parity reflection $y \to -y$. The Standard Model (SM) fields are confined to a 3-brane (with 4D world-volume) located at $y = 0$, while gravity spills into the 5D bulk~\cite{Antoniadis:1998ig}.  The dark dimension can also be viewed as a line interval with end-of-the-world 9-branes (EW9-branes) attached at each end~\cite{Schwarz:2024tet}. 

The 4D effective field theory of this geometry naturally reveals a dense, light tower of massive Kaluza-Klein (KK) graviton excitations. Their mass spacing is roughly $m_{\rm KK} \sim 0.1~{\rm eV}$, a scale determined directly by the micron-sized radius $R$ of the dark dimension. Behaving entirely as weakly interacting and collisionless particles, these {\it dark gravitons} present an elegant candidate for DM~\cite{Gonzalo:2022jac}. They are produced in the early universe through gravitational coupling to thermal matter on the brane, yielding KK modes with masses $m_n = n m_{\rm KK}$ for each index $n$. Unlike typical frozen DM profiles, dark gravitons realize a dynamical DM framework through an intra-tower cascade, where heavier KK modes decay into lighter ones~\cite{Dienes:2011ja}.
 
To evade radiative decay signatures~\cite{Law-Smith:2023czn} and prevent the premature disruption of bound structures~\cite{Obied:2023clp}, a dynamical dark sector governed by KK tower cascades demands a rapidly evolving tower with a time-dependent mass scale of $m_n(t) \propto t^{-2/7}$. While restricting the 5D effective theory to two derivatives can suppress the KK graviton cascade rate, slowing the mass evolution to $m_n(t) \propto t^{-2/3}$~\cite{Lee:2026myr,Langhoff:2026wiz}, this can be bypassed by introducing higher-derivative terms. Alternatively, a more natural solution for connection with DESI findings allows the dark gravitons to decay promptly into the KK tower of a quintessence field; the resulting scalar tower cascade preserves the $t^{-2/7}$ scaling, ensuring that the cosmological framework of Ref.~\cite{Gonzalo:2022jac} remains valid~\cite{Vafa:2026ccm}.

A compelling feature of this framework is the exponential coupling between DE and DM ($m_{\rm c}(\phi)\propto e^{-\beta\phi/M_{\rm Pl}}$), which is driven by a single scalar field that dictates the size of the dark dimension~\cite{Bedroya:2025fwh}. As the scalar field rolls over cosmic time, the properties of the dark sector do not remain static but instead evolve together at an exponential rate. Because DM in this scenario is composed of a tower of heavy particles native to the extra dimension, its mass scale is inextricably tied to the dimension's radius. As the field shifts, the exponential relationship ensures a continuous, proportional transfer of energy between DM and DE. By establishing that DM particles are slowly shifting in mass as DE changes, this dynamic exponential coupling perfectly mimics a phantom DE signature without violating fundamental quantum gravity constraints.

The framework elegantly accommodates this dynamic behavior through a geometric stabilization mechanism where the dark dimension's radius varies over cosmic time while $M_{\rm Pl}$ remains constant, a process analogous to morphing a rugby ball into a cigar. In string theory compactifications, our universe's gravitational strength depends entirely on the total volume of all hidden extra dimensions combined. By deforming the internal geometry, stretching one dimension while symmetrically squeezing another to preserve the total volume, the radius of the dark dimension can shift independently. This shape-shifting behavior redshifts the KK DM particles inhabiting the dark dimension, leading to a decreasing effective mass from the 4D perspective; yet because the overall higher-dimensional volume is conserved during this transition, $M_{\rm Pl}$ stays locked.

In addition, the dark dimension scenario elegantly explains tiny neutrino masses~\cite{Dienes:1998sb,Arkani-Hamed:1998wuz,Dvali:1999cn,Montero:2025hye}. Because right-handed neutrinos could propagate through the 5D bulk, the coupling between these bulk states and the brane-localized, left-handed SM neutrinos scales inversely with the square root of the bulk volume. Furthermore, if a gauged $B-L$ symmetry is introduced into this framework, it must act as a bulk gauge symmetry to preserve gauge anomaly cancellation via charged bulk modes. This arrangement naturally mandates the existence of three right-handed bulk neutrinos~\cite{Montero:2025hye}. When a $B-L$-charged bulk scalar triggers spontaneous symmetry breaking, it generates a massive gauge boson with an exceptionally feeble coupling ($g_{B-L} \sim 10^{-10}$) and forces these bulk states to acquire a heavy Majorana mass, dynamically generating small active neutrino masses via the seesaw mechanism. To avoid the catastrophic overproduction of KK neutrino towers in the early universe~\cite{Abazajian:2000hw}, the masses of these bulk Majorana neutrinos must satisfy a strict cosmological lower bound of approximately $100~{\rm GeV}$~\cite{Anchordoqui:2026ret}. Consequently, the effective number of relativistic degrees of freedom predicted by the model remains tightly constrained at the SM value of $N_{\rm eff} = 3.044$, ensuring perfect consistency with precision early universe cosmology and recent CMB observations.

\section{Alternative Supernova Samples}
\label{app:SN}

We further assess the dependence on the Type Ia supernova compilation by replacing the baseline DES-Dovekie data with PantheonPlus~\cite{Brout:2022vxf} or Union3~\cite{Rubin:2023jdq}. The resulting constraints, presented in Table~\ref{tab:SN_results}, show that the CQ neutrino-mass constraints remain stable across the supernova samples, while CPL exhibits larger shifts in both the effective neutrino mass and the DE parameters. The CQ and CPL neutrino mass constraints become closer with PantheonPlus, whereas Union3 shifts the CPL posterior mean toward more positive values, leaving CQ nearly unchanged. The constraints on the CQ coupling also remain stable around $\beta\simeq0.04$ across all three supernova compilations.

\begin{table}[htbp]
\centering
\setlength{\tabcolsep}{2mm}
\renewcommand{\arraystretch}{1.4}
\scriptsize
\resizebox{0.95\columnwidth}{!}{%
\begin{tabular}{lcc}
\hline
\hline
Model/Parameters &
{CMB+DESI+PantheonPlus} &
{CMB+DESI+Union3} \\
\hline
$\bm{\Lambda}$\textbf{CDM}+$\sum m_{\nu,\mathrm{eff}}$ & & \\
$\sum m_{\nu,\mathrm{eff}}$~[eV] & $-0.044^{+0.032}_{-0.041}$ & $-0.048^{+0.033}_{-0.038}$ \\
$H_0\ [{\rm km\,s^{-1}\,Mpc^{-1}}]$ & $68.66\pm0.32$ & $68.70\pm0.31$ \\
$\Omega_{\mathrm{m}}$ & $0.2989\pm0.0038$ & $0.2985\pm0.0037$ \\
$\sigma_8$ & $0.8327\pm0.0083$ & $0.8335\pm0.0080$ \\
$S_8$ & $0.8312\pm0.0078$ & $0.8313\pm0.0077$ \\
\hline
$\bm{\textbf{CPL}}$+$\sum m_{\nu,\mathrm{eff}}$ & & \\
$\sum m_{\nu,\mathrm{eff}}$~[eV] & $-0.016^{+0.042}_{-0.054}$ & $0.002^{+0.046}_{-0.061}$ \\
$w_0$ & $-0.858\pm0.053$ & $-0.696\pm0.089$ \\
$w_a$ & $-0.47^{+0.22}_{-0.19}$ & $-0.92^{+0.33}_{-0.28}$ \\
$H_0\ [{\rm km\,s^{-1}\,Mpc^{-1}}]$ & $67.71\pm0.59$ & $66.17\pm0.84$ \\
$\Omega_{\mathrm{m}}$ & $0.3087\pm0.0057$ & $0.3243\pm0.0088$ \\
$\sigma_8$ & $0.8237\pm0.0093$ & $0.810\pm0.011$ \\
$S_8$ & $0.8354\pm0.0079$ & $0.8415\pm0.0083$ \\
\hline
$\bm{\textbf{CQ}}$+$\sum m_{\nu,\mathrm{eff}}$ & & \\
$\sum m_{\nu,\mathrm{eff}}$~[eV] & $-0.016^{+0.040}_{-0.064}$ & $-0.017^{+0.039}_{-0.066}$ \\
$\beta$ & $0.039^{+0.017}_{-0.012}$ & $0.042^{+0.015}_{-0.012}$ \\
$\alpha$ & $0.35^{+0.17}_{-0.21}$ & $0.47\pm0.21$ \\
$H_0\ [{\rm km\,s^{-1}\,Mpc^{-1}}]$ & $68.08\pm0.60$ & $67.64\pm0.70$ \\
$\Omega_{\mathrm{m}}$ & $0.2992\pm0.0053$ & $0.3018\pm0.0058$ \\
$\sigma_8$ & $0.8323\pm0.0095$ & $0.8293\pm0.0094$ \\
$S_8$ & $0.8311\pm0.0079$ & $0.8317\pm0.0077$ \\
\hline
\hline
\end{tabular}%
}
\caption{ \label{tab:SN_results} Constraints at the 68\% CL in the
$\Lambda$CDM+$\sum m_{\nu,\mathrm{eff}}$, CPL+$\sum m_{\nu,\mathrm{eff}}$,
and CQ+$\sum m_{\nu,\mathrm{eff}}$ models.}
\end{table}

\section{Additional Results for Phenomenological IDE Models}
\label{app:IDE}

\begin{figure}[t!]
\centering
\includegraphics[width=0.47\textwidth]{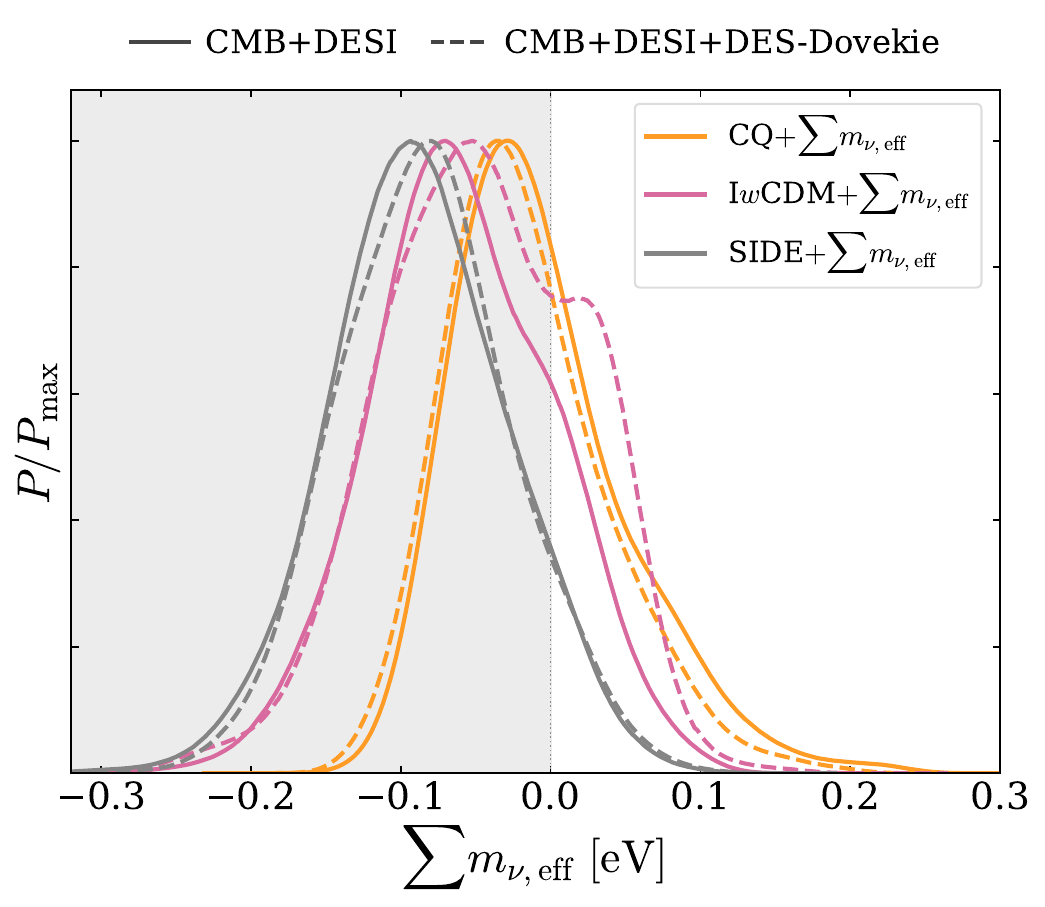}
\caption{Marginalized posterior distributions of $\sum m_{\nu,\mathrm{eff}}$ for the CQ+$\sum m_{\nu,\mathrm{eff}}$, I$w$CDM+$\sum m_{\nu,\mathrm{eff}}$, and SIDE+$\sum m_{\nu,\mathrm{eff}}$  models, obtained from CMB+DESI and CMB+DESI+DES-Dovekie data.}
\label{fig:IDE_mnu:posteriors}
\end{figure}

\begin{table*}[htbp!] 
\centering
\setlength{\tabcolsep}{2mm}
\renewcommand{\arraystretch}{1.4}
\scriptsize
\begin{tabular}{lcccc}
\hline
\hline
Model/Parameters &
{CMB+DESI} &
{CMB+DESI+DES-Dovekie} &
{CMB+DESI+PantheonPlus} &
{CMB+DESI+Union3} \\
\hline
$\bm{\textbf{I}w\textbf{CDM}}+\sum m_{\nu,\mathrm{eff}}$ & & & & \\
$\beta$ & $-1.36\pm0.62$ & $-1.70^{+0.92}_{-0.39}$ & $-1.55^{+0.74}_{-0.67}$ & $-1.83^{+0.73}_{-0.62}$ \\
$w$ & $-1.23^{+0.14}_{-0.12}$ & $-1.32^{+0.21}_{-0.09}$ & $-1.28\pm0.15$ & $-1.30\pm0.14$ \\
$\sum m_{\nu,\mathrm{eff}}$~[eV] & $-0.064\pm0.062$ & $-0.039\pm0.066$ & $-0.050\pm0.068$ & $-0.063\pm0.066$ \\
$H_0\ [{\rm km\,s^{-1}\,Mpc^{-1}}]$ & $67.50\pm1.20$ & $67.66^{+0.49}_{-0.59}$ & $67.70\pm0.61$ & $66.90\pm0.80$ \\
$\Omega_{\mathrm{m}}$ & $0.502^{+0.092}_{-0.071}$ & $0.534^{+0.065}_{-0.085}$ & $0.520^{+0.082}_{-0.068}$ & $0.557^{+0.071}_{-0.063}$ \\
$\sigma_8$ & $0.568^{+0.037}_{-0.085}$ & $0.538\pm0.055$ & $0.550^{+0.039}_{-0.076}$ & $0.525^{+0.032}_{-0.056}$ \\
$S_8$ & $0.725^{+0.016}_{-0.035}$ & $0.711\pm0.026$ & $0.716^{+0.020}_{-0.035}$ & $0.710^{+0.015}_{-0.025}$ \\
\hline
$\bm{\textbf{SIDE}}+\sum m_{\nu,\mathrm{eff}}$ & & & & \\
$\beta_0$ & $<-0.56$ & $-1.27^{+0.80}_{-0.58}$ & $-0.77^{+0.72}_{-0.59}$ & $-2.40\pm1.00$ \\
$\beta_e$ & $2.9^{+2.1}_{-1.6}$ & $1.5^{+1.2}_{-1.4}$ & $0.6^{+1.1}_{-1.4}$ & $2.7\pm1.6$ \\
$\sum m_{\nu,\mathrm{eff}}$~[eV] & $-0.087\pm0.060$ & $-0.084^{+0.056}_{-0.062}$ & $-0.102\pm0.061$ & $-0.075\pm0.060$ \\
$H_0\ [{\rm km\,s^{-1}\,Mpc^{-1}}]$ & $65.60\pm1.20$ & $67.45\pm0.55$ & $67.65\pm0.60$ & $66.17\pm0.83$ \\
$\Omega_{\mathrm{m}}$ & $0.564^{+0.110}_{-0.085}$ & $0.429^{+0.046}_{-0.056}$ & $0.391\pm0.051$ & $0.528^{+0.078}_{-0.066}$ \\
$\sigma_8$ & $0.512^{+0.030}_{-0.092}$ & $0.619^{+0.059}_{-0.077}$ & $0.678^{+0.063}_{-0.101}$ & $0.528^{+0.038}_{-0.077}$ \\
$S_8$ & $0.688^{+0.016}_{-0.049}$ & $0.734^{+0.035}_{-0.042}$ & $0.766^{+0.038}_{-0.052}$ & $0.693^{+0.024}_{-0.044}$ \\
\hline
\hline
\end{tabular}
\caption{\label{tab:ide_models} Constraints at the 68\% CL and upper limits at 95\% CL in the I$w$CDM+$\sum m_{\nu,\mathrm{eff}}$ and
SIDE+$\sum m_{\nu,\mathrm{eff}}$ models.}
\end{table*}

In addition to the general theory of CQ, we also consider two phenomenological interacting DE models with two additional parameters as complementary scenarios. These models modify the evolution of the DM and perturbations in different ways, allowing us to test how the interplay between energy transfer, expansion, and structure growth affects neutrino mass constraints.

The first model, referred to as I$w$CDM, treats DE as a fluid with a constant EoS parameter $w\neq -1$~\cite{Giare:2024ytc,Li:2026jgg}. The interaction is specified phenomenologically as~\cite{Li:2024qso,Li:2026jgg}
\begin{equation}
    Q=\beta H_0\rho_{\rm de},
\end{equation}
where $\beta$ is a dimensionless coupling parameter. Thus, I$w$CDM extends the $\Lambda$CDM model by introducing two additional parameters, $w$ and $\beta$. The direction of energy transfer is determined by the sign of the coupling $\beta$: DM decays into DE for $\beta>0$, and vice versa for $\beta<0$. In our analysis, $\beta$ can take both positive and negative values.

The second is a sign-changeable interacting DE model, referred to as SIDE. In this case, the direction of energy transfer may change during cosmic evolution~\cite{Cai:2009ht,Li:2011ga,Li:2025owk}. We fix the DE EoS parameter to $w=-1$ and consider the interaction term $Q=\beta(a)H_0\rho_{\rm de}$. The time dependence of the coupling is parametrized as~\cite{Li:2011ga}
\begin{equation}
\beta(a)=\beta_{0}a+\beta_{\rm e}(1-a),
\end{equation}
where $\beta_{0}$ and $\beta_{\rm e}$ are dimensionless parameters that characterize the late- and early-time coupling strengths, respectively. SIDE also extends the $\Lambda$CDM model by introducing the two additional parameters, $\beta_{0}$ and $\beta_{\rm e}$. A zero crossing of $\beta(a)$ indicates a reversal in the direction of energy transfer between DE and DM~\cite{Li:2025owk,Li:2025ula}.

The resulting constraints are summarized in Table~\ref{tab:ide_models}, while Fig.~\ref{fig:IDE_mnu:posteriors} compares the effective neutrino mass posteriors with those obtained in CQ. For the two baseline data combinations, both models yield more negative posterior means than CQ. In I$w$CDM, adding DES-Dovekie shifts the mean from $-0.064\,\mathrm{eV}$ to $-0.039\,\mathrm{eV}$, whereas SIDE remains centered at more negative values, $-0.087\,\mathrm{eV}$ and $-0.084\,\mathrm{eV}$, respectively. The same ordering of the posterior means across the three models persists when using PantheonPlus or Union3.

Interestingly, both models favor substantially larger values of $\Omega_m$ and lower values of $\sigma_8$ and $S_8$ than CQ. In I$w$CDM, the preferred negative coupling corresponds to energy transfer from DE to DM and is associated with a phantom DE equation of state. SIDE instead favors negative late-time coupling values, with a greater dependence of the early-time coupling on the supernova sample. Despite their lower clustering amplitudes, neither model shifts the effective neutrino-mass posterior toward positive values as much as CQ. These results show that increasing $\Omega_m$ or lowering $S_8$ (or $\sigma_8$) alone is not sufficient to weaken neutrino-mass bounds. Larger masses can be accommodated only when the changes induced in the expansion history and structure growth work together to reconcile the early-time CMB constraints with late-time distances and lensing.

\end{document}